\documentclass[a4paper]{article}
\usepackage[dvips]{epsfig}
\newcommand{\tfrac}{\frac}
\newcommand{\dfrac}{\frac}
\addtocounter{secnumdepth}{1}
\newcommand{\dd}{\text{d}}
\newcommand{\ee}{\text{e}}
\newcommand{\ii}{\text{i}}
\newcommand{\ji}{\text{\small i}}
\newcommand{\p}{\partial}

\newcommand{\Ai}{\text{Ai}}
\newcommand{\text}{\mbox}       

\def\gsim{\:\raisebox{-0.5ex}{$\stackrel{\textstyle>}{\sim}$}\:}
\newcounter{exerc}
\begin{document}

\thispagestyle{empty}
\title{Persistence exponents\\ 
of non-Gaussian processes\\
in statistical mechanics}
 
\author{O. Deloubri{\`e}re and H.J. Hilhorst\\
Laboratoire de Physique Th{\'e}orique$^1$\\
B{\^a}timent 210\\
Universit{\'e} de Paris-Sud\\
91405 Orsay cedex, France\\}

\maketitle
\vspace{-1cm}
\begin{small}
\begin{abstract}
\noindent 
Motivated by certain problems of statistical physics 
we consider  a stationary stochastic process in which
deterministic evolution is interrupted at random times by upward jumps 
of a fixed size. If the evolution consists of
linear decay, the sample functions are of the "random sawtooth" type
and the level dependent persistence
exponent $\theta$
can be calculated exactly. We then develop an expansion method
valid for small curvature of the deterministic curve. The curvature
parameter $g$   
plays the role of the coupling constant of an interacting particle
system. 
The leading order curvature correction to $\theta$
is proportional to $\sim g^{2/3}$. 
The expansion applies in particular to exponential decay in the limit of
large level, where the curvature correction 
considerably improves the linear approximation.
The Langevin equation, with Gaussian white noise, is recovered
as a singular limiting case. 
\end{abstract}
\end{small}
\vspace{90mm}

\noindent L.P.T. -- ORSAY 99/70\\
{\small$^1$Laboratoire associ{\'e} au Centre National de la
Recherche Scientifique - UMR 8627}
\newpage
\section{Introduction} 
\label{secintroduction}

In this work we study the stationary
stochastic process $\xi(t)$ that obeys the equation 
\begin{equation}
\frac{\dd \xi(t)}{\dd t}=- A(\xi)+a\,\sum_\ell\delta(t-t_\ell)
\label{defxi}
\end{equation}
Here $a$ is a positive parameter and the $t_\ell$
are random times distributed independently and uniformly with density
$\rho$; the random term therefore represents white noise, but with
a nonzero average equal to $\rho a$. 
Hence $\xi(t)$ evolves deterministically except for upward jumps of
fixed size $a$ occurring at random times. 
We take the systematic "force"
$A(\xi)$ such that it has positive
derivative and satisfies $A(-\infty)\!<\!\rho a\!<\!A(\infty)$,
which ensures that $\xi$ possesses a stationary distribution.
A special case is the linear equation obtained for the 
choice $A(\xi)=\beta\xi$. 
Our interest is in 
the first passage time problem associated with 
a pre\-established threshold $\xi=X$.

More precisely, for some general stationary process $\xi(t)$,
let $Q(T)$ be the probability that during a time interval 
of length $T$ it stays above a threshold $X$,
given that it was larger than
$X$ at the beginning of that interval. 
For many of the common processes in physics 
$Q(T)$ decays to zero exponentially with an
{\it inverse relaxation time} $\theta$ defined by
\begin{equation}
\theta=\lim_{T\to\infty}T^{-1}\log Q(T)
\label{deftheta}
\end{equation}
Both $Q(T)$ and $\theta$ depend on the threshold value $X$.\\

Physicists are interested in the persistence exponents of various
stochastic processes because of their connection to
critical phenomena; there, after an appropriate rescaling of
variables, $\theta$
appears as the exponent of a power law 
and is called the {\it persistence exponent}.
The theory of critical phenomena has brought to
light the importance of exponents for the classification of physical
systems.
This has spurred theoretical physicists
in recent years
to attempt to calculate persistence exponents associated with 
several prominent problems of that discipline.
The exponents are
in each case nontrivial and unrelated to the 
static and dynamic critical exponents of the same problem.

Many authors have studied processes of zero average and symmetric under
sign change of $\xi$. The
quantity of primary
interest is then "the" persistence exponent associated with the
threshold $X=0$. For nonzero $X$ one also speaks of the {\it level} exponents.
A review article of earlier work, mainly mathematical, is due to Blake and
Lindsay \cite{BlakeLindsay}. 
Majumdar \cite{Majumdar} and Godr{\`e}che \cite{Godreche}
have provided useful reviews of recent work,
mainly by physicists.
Almost all of this work
deals with processes $\xi(t)$ that are Gaussian. Among these the
Gaussian Markovian case is easiest to treat. 
Majumdar and Sire \cite{MajumdarSire}, followed by Oerding {\it et al.}
\cite{OCB}  and
Sire {\it et al.} \cite{SMR}, have designed a perturbative
method for processes that are Gaussian and close to Markovian.
Majumdar and Bray \cite{MajumdarBray} 
have set up an $\varepsilon$ expansion for smooth
Gaussian processes in spatial dimension $d=4-\varepsilon$.
Nontrivial persistence exponents have also been identified 
for such familiar functions as the
solution of the diffusion equation with random initial condition
\cite{MSBC,DHZ}.
All these persistence exponents appear to be
extremely difficult to calculate analytically.\\

In physical systems the addition of the effects of
many degrees of freedom 
very often leads to Gaussian processes. Nevertheless, certain
non-Gaussian processes also arise naturally.
In this work we study the level exponents for the strongly
non-Gaussian case of Eq.\,(\ref{defxi}).
One example of how closely related processes enter
physics is through a question
\cite{GDH} associated with the one-dimensional random walk. Let 
$\tilde{\xi}(\tilde{t})$ 
be the number of steps needed before the walk has visited $\tilde{t}$
distinct sites.
Then $\xi(t)\equiv\ee^{-t}\,\tilde{\xi}(\ee^{2t})$
is a stationary process consisting of exponential
decay interrupted by upward jumps.
The only difference with Eq.\,(\ref{defxi})
is that it leads to a probability distribution of
the jump sizes, whereas in (\ref{defxi}) we take $a$ a fixed parameter.
Numerical evidence shows that this problem possesses
well-defined persistence
exponents, but, as in many other cases, no way
exists to find them analytically.

Other instances are furnished by statistical physical models
that exhibit
cluster growth, such as
Random Sequential Absorption and percolation theory; if $\xi(t)$ is the
suitably scaled
size of a particular cluster, then jumps are due to coalescence with
other clusters.  As a specific example, let the bonds of a
lattice be rendered percolating in a sequential manner
\cite{FreitasLucena} and let
$\tilde{t}$ be the instantaneous fraction of
percolating bonds; define then
$\tilde{\xi}(\tilde{t})$ as the size of the cluster connected to the
origin. In spatial dimension one it is easily shown \cite{Deloubriere}
that an appropriate scaling (which is such that $t\to\infty$
as $\tilde{t}\to 1$) yields again a stationary process $\xi(t)$ with a
probablity law for the jump sizes.\\

In order to study the persistence exponents associated with Eq.\,(\ref{defxi})
we exploit the following idea.
The persistence probability $Q(T)$ is determined by the subclass
of $\xi(t)$ that do not cross below $X$ for $0<t<T$.
When the force $A(\xi)$ 
is strongly positive, then 
a $\xi(t)$ in the contributing subclass
is unlikely ever to rise very high above the threshold $X$. 
We conjecture, therefore, that we will
obtain a good description
of this subclass by expanding
$A(\xi)$ around the threshold value $\xi=X$.
We convert this idea into an expansion procedure.
Roughly speaking, the zeroth and the first
order of the expansion are determined by
$A(X)$ and $A'(X)/A(X)$, respectively, {\it i.e.,} by the slope and the
curvature/(slope)$^2$ ratio of the deterministic evolution curve.
The precise mathematics is slightly more subtle and
shows that instead two parameters appear,
called $r$ and $g$, whose definition is more complicated.
Our theory then yields the level exponents $\theta$ in terms of
$r$ and $g$ in the small $g$ limit. We shall refer to $g$ as the {\it
curvature parameter.}\\

In Section \ref{secgeneralities} we write $Q(T)$ as a path integral on
all contributing $\xi(t)$.
The expression resembles the partition
function of a system of interacting particles in a 
one-dimensional volume $T$, with
the jump times $t_1,t_2,\ldots$ in the role of the particle positions.
We rearrange the path integral in such a way that a "noninteracting"
contribution appears, characterized by a parameter $r$, and a remainder
due to an
"interaction potential" $V$ which is a functional of the jump times.
Our use of the term "noninteracting" does not mean that the $V=0$
problem is
trivial -- it is not --, but merely that it is purely combinatorial.
We are led to define the parameter $r$ of the noninteracting theory by
\begin{eqnarray}
r&=&\rho\,\int_X^{X+a}\frac{\dd\xi}{A(\xi)}\nonumber\\[2mm]
&=&\frac{\rho a}{A(X)}\,\Big(1\,-\,\frac{aA'(X)}{2A(X)}+\ldots\Big)
\label{defrA}
\end{eqnarray}
This equation shows that $r$ involves not only
$A(X)$ but also the full series of its derivatives.

In Section \ref{seclinear} we consider the zeroth order,
$V=0$. 
It amounts to replacing
$A(\xi)$ in Eq.\,(\ref{defxi}) by the constant $\rho a/r$, so that as
a consequence $\xi(t)$ is piecewise linear with slope $-\rho a/r$. 
All samples of this zeroth order
process are therefore "random sawtooth" functions.
In this order
we shall write the level exponent as $\theta_{0}(X)$.
We find
\begin{equation}
\theta_{0}(r{})=\rho\,\Big(
\frac{1}{r{}}\log\frac{1}{r{}}-\frac{1}{r{}}+1\Big) \qquad 0<r<1
\label{resthetalin}
\end{equation}
For $r{}\to 1$ the
persistence exponent goes to zero; the interpretation of this unphysical
effect is that
for $r{}>1$ the linearization creates
a finite probability for $\xi(t)$ to escape
to $+\infty$.\\

In Section \ref{secnonlinear} we consider the interacting theory,
\,$V\neq 0$. 
The potential $V$ is determined by $A$ in a way described in that
section. We are unable to deal with the general case.
Instead, we expand $V$ in a series of which we retain only the first
term, whose coefficient $g$ plays the role of an interaction constant. 
The expression for $g$ is 
\begin{eqnarray}
g&=&\frac{A(X+a)-A(X)}{A(X)}\nonumber\\[2mm]
&=&\frac{aA'(X)}{A(X)}\,
\Big(1\,+\,\frac{aA''(X)}{A(X)}\,+\ldots\Big)
\label{defgA}
\end{eqnarray}
We show that there are at least two limits in which the 
higher order terms in the series for $V$ 
are negligible, and in which the remaining problem, with only
two parameters $r$ and $g$, can be solved.

The Subsections \ref{secrecursionKom} and
\ref{secanalysisKOm} are common to both limits. The Laplace
transform $K(\Omega)$ of the
path integral for $Q(T)$ appears to satisfy a
recursion relation whose solution is expressed in
Eq.\,(\ref{solKOm}) as the ratio of
two infinite series. Explicit
evaluation of these series turns out to be a rather formidable task.
The soluble limits are the following.

{\it Limit (i).} The limit $g\to 0$ at $r$ fixed.
Eqs.\,(\ref{defrA}) and (\ref{defgA}) show that this corresponds to $A'(X)\to 0$
at fixed $A(X)$.
In Subsection \ref{secdelto0}
we calculate the exponent $\theta(r{},g{})$ in a small
$g{}$ expansion, with the result that a nonanalytic correction term to
Eq.\,(\ref{resthetalin}) appears,
\begin{equation}
\theta(r,g)=\theta_{0}(r{})+
\rho\,\frac{1}{2r{}}\Big (\log\frac{1}{r{}}\Big )^{2/3}\,
\Big (\frac{9\pi}{4}g{}\Big )^{2/3}
\qquad g{}\to 0, \quad r \mbox{ fixed } 
\label{restaudelta}
\end{equation}

{\it Limit (ii).} The limit $g\to 0,\, r\to 0$ with fixed ratio
$g/r$. Eqs.\,(\ref{defrA}) and (\ref{defgA}) show that this corresponds to
$A(X)\to\infty\,$ at fixed $A'(X)$. 
In this limit the expansions in Eqs.\,(\ref{defrA}) and (\ref{defgA})
may be replaced by their first term, which we shall denote by an index 0,
\begin{equation}
r_0=\frac{\rho a}{A(X)}, 
\qquad g_0=\frac{aA'(X)}{A(X)}
\label{defaldel}
\end{equation}
This limit, considered in Subsection \ref{secaldelto0},
requires separate analysis; nevertheless, the result for
$\theta(r,g)$ is 
what one also obtains by naively substituting 
$r=r_0$ and $g=g_0$ in 
Eq.\,(\ref{restaudelta}). 

The example of greatest interest is the
linear equation that prevails for the choice $A(\xi)=\beta\xi$. When the
threshold $X$ becomes large we have $r_0=\rho a/\beta X$ and
$g_0=a/X$. 
Upon expressing for this case $\theta$ as a 
function of $X$ we arrive at
the explicit asymptotic expansion
\begin{equation}
\theta(X)\,=\,\theta_{0}\Big(\frac{\rho a}{\beta X}\Big)
\,+\,\frac{\beta}{2}\Big(\frac{9\pi}{4}\Big)^{2/3}
\Big(\frac{\beta X}{\rho a}\Big)^{1/3}
\Big (\log\frac{\beta X}{\rho a}\Big )^{2/3}\,+\,\ldots    \qquad X\to\infty
\label{resthetaexp}
\end{equation}
In Section \ref{secapplications}
we compare analytical results for both limit cases to
Monte Carlo simulations of Eq.\,(\ref{defxi}).
Excellent agreement is found. In particular, there is strong numerical 
indication that the higher order terms in the asymptotic
expansion (\ref{resthetaexp}) go to zero as $X\to\infty$.\\

The name {\it Langevin equation} 
is traditionally reserved for equations of type (\ref{defxi}) 
where the
random term represents Gaussian white noise. 
In Section \ref{seccomparison}
we observe that 
the white noise of Eq.\,(\ref{defxi}) becomes Gaussian 
in the limit $\rho\to\infty$ and $a\to 0$ at fixed $\rho a^2$, and that,
correspondingly (and 
after appropriate rescaling of variables) Eq.\,(\ref{defxi}) becomes a
Langevin equation.
Hence our work enables us to pass continuously from strongly
non-Gaussian to Gaussian noise.
In Subsection \ref{secthetaG} we place ourselves directly in the
Gaussian limit and determine,
{\it via} the associated Fokker--Planck equation, the
Gaussian persistence 
exponent $\theta_G$ for asymptotically high threshold; 
our method is close to the one
of Krapivsky and Redner \cite{KrapivskyRedner}. 
In Subsection \ref{secglimit}
we then investigate how the
Gaussian limit emerges from the more general approach of 
Sections \ref{secgeneralities}--\ref{secapplications}.

Section \ref{secconclusion} contains our conclusions.

\section{Phase space integral}
\label{secgeneralities}

\subsection{Solution $\xi(t)$}
\label{secsolution}

The solution of Eq.\,(\ref{defxi}) is piecewise continuous.
In the time interval between two jumps $\xi(t)$ 
evolves deterministically  according to
\begin{equation}
\xi(t)=f(t-u_\ell), \quad\qquad t_{\ell-1}<t<t_\ell
\label{interval}
\end{equation}
where $u_\ell$ determines a shift along the time axis and
the function $f(t)$, if we choose it such that $f(0)=X$, is obtained from $A(\xi)$ by 
\begin{equation}
t=-\int_{X}^{f(t)}\frac{\dd\xi}{A(\xi)}
\label{deff}
\end{equation}
Hence $u_\ell$ acquires the meaning of the ultimate instant of time at which
the $\ell$th jump should take place if $\xi(t)$ is to stay above the
threshold.
The fact that there is a jump of
size $a$ on the border between two successive time
intervals leads to the identity
\begin{equation}
f(t_\ell-u_{\ell+1})-f(t_\ell-u_\ell)=a 
\label{f4}
\end{equation}

We shall be more specific now and consider the solution $\xi(t)$ of
Eq.\,(\ref{defxi}) with initial value
$\xi(0)=\xi_0$.
It is uniquely specified by the set
of jump times $0<t_1<t_2<\ldots .$
Eq.\,(\ref{f4}), which is here valid for $\ell=1,2,\ldots,$ 
allows one to express
$u_{\ell+1}$ in terms of $t_{\ell}$ and $u_{\ell}$, and, upon
iterating, as a function of $t_1,\ldots,t_\ell$ and $u_1$. Finally,
$u_1$  may be eliminated in favor of the initial value $\xi_0$ by means
of $f(-u_1)=\xi_0$.
Hence we have obtained
the formal answer to the question of how to find 
$u_\ell$ as a function of the random jump times and the initial
condition. Below it will be convenient to use $t_0\equiv 0$ and 
$u_0\equiv 0$; with
that convention Eq.\,(\ref{f4}) holds also for $\ell=0$ {\it if} we take
the special initial condition $\xi_0=X+a$.\\

\subsection{Basic integral}
\label{secbasic}

The persistence probability $Q(T)$ 
can be expressed as a path integral on all random functions $\xi(t)$,
hence as an integral on all jump times $t_1,t_2,\ldots .$ It is now
useful to note that
the $u_\ell$ are ordered according to $0=u_0<u_1<u_2<\ldots,$ so that
there exists an $L\geq 0$ for which
\begin{equation}
u_L<T<u_{L+1}
\label{defL}
\end{equation}
The interpretation is that after the $L$th jump the function $\xi(t)$ is
sure to stay above the threshold $X$, even if no further jumps occur, in
the interval $[0,T]$.
Summing on all possibilities implied by Eq.\,(\ref{defL}) 
we can write $Q(T)$ as
\begin{equation}
Q(T)=\sum_{L=0}^\infty\rho^L
\int_0^{u_1}\!\dd t_{1}
\int_{t_{1}}^{u_2}\!\dd t_{2}\ldots\int_{t_{L-1}}^{u_L}\!\dd t_{L}
\,\ee^{-\rho t_L}\,\Theta(T-u_L)\Theta(u_{L+1}-T)
\label{sumL}
\end{equation}
where $\Theta$ is the Heaviside step function and where we
used that $\rho^L\,\ee^{-\rho t_L}$ is the joint probability density for
the first $L$ jumps to occur at $t_1,t_2,\ldots,t_L$. The $L=0$ term in
Eq.\,(\ref{sumL}) has no integrals and is equal to $\Theta(u_1-T)$. 
In the remainder we will use the shorthand notation
\begin{equation}
\int_0=1, \qquad \int_{\ell}=\rho^\ell\int_0^{u_1}\!\dd t_{1}
\int_{t_{1}}^{u_2}\!\dd t_{2}\ldots\int_{t_{\ell-1}}^{u_\ell}\!\dd t_{\ell}
\qquad \ell=1,2,\ldots
\label{defintL}
\end{equation} 
The expression (\ref{sumL})
for $Q(T)$ bears great similarity to the grand-canonical
partition function of an assembly of interacting particles in a
one-dimensional volume $T$, with the jump times $t_1,t_2,\ldots$ playing
the role of the particle positions and with the interaction implicit in the
upper integration limits $u_1,u_2,\ldots.$

In terms of Laplace transforms Eq.\,(\ref{sumL}) is equivalent to
\begin{eqnarray}
\hat{Q}(\omega)&\equiv&\int_0^\infty\dd T\,\ee^{-\omega T}Q(T)\nonumber\\
&=& \omega^{-1}\sum_{L=0}^\infty\,\int_L\ee^{-\rho t_L}
(\ee^{-\omega u_L}-\ee^{-\omega u_{L+1}})
\label{defQhat}
\end{eqnarray}
One more rewriting is useful.
For $L\geq 1$ one easily finds the relation
\begin{equation}
\int_L\ee^{-\rho t_L-\omega u_L}=
\int_{L-1}\ee^{-\rho t_{L-1}-\omega u_{L}}
-\int_{L-1}\ee^{-(\rho+\omega)u_L}
\label{aux1}
\end{equation}
When Eq.\,(\ref{aux1}) is substituted in Eq.\,(\ref{defQhat})
cancellations occur. After we replace $\omega$ with the
dimensionless variable
\begin{equation}
\Omega=\frac{\rho+\omega}{\rho}
\label{defOm}
\end{equation}
we can express the problem by the three equations
\begin{eqnarray}
\omega\hat{Q}(\omega)&=&1-{K}(\Omega)\label{QoKo}\\
{K}(\Omega)&=&\sum_{L=0}^\infty K_L(\Omega)\label{KoKL}\\
K_L(\Omega)&=&\int_L\ee^{-\Omega\rho u_{L+1}} \qquad L=0,1,\ldots
\label{finalgen}
\end{eqnarray}
of which the last one implies, in particular, that
$K_0(\Omega)=\ee^{-\Omega\rho u_1}$.
Our task is to evaluate the phase space integral $\int_L$
in Eq.\,(\ref{finalgen}) and to find the relevant nonanalyticity of
$\hat{Q}(\omega)$. 
In terms of the Laplace variable $\omega$
the persistence exponent $\theta$ will be given by
\begin{equation}
\theta=-\omega_1=-\rho(\Omega_1-1)
\label{relthetaom1}
\end{equation}
where $\omega_1$ is the real part of the rightmost nonanalyticity of
$\hat{Q}(\omega)$ in the complex $\omega$ plane,
and $\Omega_1$ is the corresponding value of $\Omega$.
Any further nonanalyticities at $\Omega_2, \Omega_3,\ldots$ will
similarly give rise to correction terms in the decay of $Q(T)$
characterized by $\theta_2, \theta_3,\ldots.$

\subsection{Interaction potential $V(y)$}
\label{secsmallcurvature}

At this stage the problem is to calculate $K_L$ of
Eq.\,(\ref{finalgen}), defined as an integral via Eq.\,(\ref{defintL}), in
which the upper integration limits $u_\ell$ are defined recursively via
Eq.\,(\ref{f4}). This problem depends 
parametrically on the function $A(\xi)$ or,
equivalently, on $f(t)$, and on the threshold $X$. We can still gain by transforming
to another set of parameters. That will be the purpose of this subsection.

Each jump provides the process with an additional lapse of time
before hitting the threshold. The extra time furnished by
the $\ell$th jump is $u_{\ell+1}-u_\ell$.
The negative slope of $f$ restricts
$u_{\ell+1}-u_\ell$ to a maximum value that we shall call $\tau$
and which occurs for $t_\ell=u_\ell$. Using this in Eq.\,(\ref{f4})  
we see that $\tau$ is the solution of
\begin{equation}
f(-\tau)-f(0)=a
\label{deftau}
\end{equation}
where, of course, $f(0)=X$.
The $\ell$th jump will generally take
place {\it before} rather than {\it at} 
the ultimate instant $u_\ell$.
Due to the upward curvature of $f$
the actual extra time gained is therefore generally less than $\tau$.
We will express this curvature effect explicitly in terms of a
variable $v_\ell$ by setting, for $\ell=1,2,\ldots,$
\begin{equation}
u_{\ell+1}-u_\ell=\tau-v_\ell
\label{defvl}
\end{equation}
whence necessarily $0<v_\ell\leq \tau$.
We now use this equation in (\ref{f4}) to eliminate $u_{\ell+1}$ and
we then subtract Eq.\,(\ref{deftau}). This gives
\begin{equation}
f(t_\ell-u_{\ell}-\tau+v_\ell)-f(t_\ell-u_\ell)
-f(-\tau)+f(0)=0
\label{eqnvl}
\end{equation}
from which $v_\ell$ can be solved in terms of
$u_\ell-t_\ell$. Although the jump density $\rho$ does not appear in the above
equation, it will turn out to be convenient to write the solution
$v_\ell$ in the scaled form
\begin{equation}
\rho\, v_\ell=V(\rho(u_\ell-t_\ell))
\end{equation} 
in which $V$ has the expansion
\begin{equation}
V(y)=\tau\sum_{k=1}^\infty g_k\,y^k
\label{defVy}
\end{equation}
It is easily seen that in accordance with Eq.\,(\ref{eqnvl}) one has
$v_\ell=0$ when \,$u_\ell-t_\ell=0$.
One obtains from Eq.\,(\ref{eqnvl}) an equation for $g_k$ in terms of $g_1,\ldots,g_{k-1}$
by differentiating $k$ times with respect to
$u_\ell-t_\ell$ and setting $u_\ell-t_\ell=v_\ell=0$. This yields
for the first two coefficients
\begin{equation}
g_1=\frac{f'(-\tau)-f'(0)}{\tau f'(-\tau)}\label{solg1}, \qquad
g_2=\frac{f''(0)}{2\rho\tau f'(-\tau)}
\,-\,\frac{f'(0)^2f''(-\tau)}{\rho\tau f'(-\tau)^3}
\label{solg2}
\end{equation} 
We emphasize that 
we do not suppose $\tau$ small.
In cases where the limit $\tau\to 0$ may be taken, obvious
simplifications occur.

We continue now the analysis of the integral (\ref{finalgen}) for $K_L.$
This analysis may be performed for general initial condition $\xi_0$;
however, from here on we shall impose $\xi_0=a$, whence $u_1=\tau,$ in
order to have simpler expressions, knowing that the persistence exponent
will not depend on $\xi_0.$ We will briefly come back to this point after
Eq.\,(\ref{exprBn}).
It is useful to define $r=\rho\tau,$ which, by Eq.\,(\ref{deftau}) and
relation (\ref{deff}) between $A$ and $f$, is equivalent to
Eq.\,(\ref{defrA}) of the Introduction.
Rewriting Eq.\,(\ref{finalgen}) in
terms of the new integration variables
$y_\ell=\rho(u_\ell-t_\ell)$
and using Eq.\,(\ref{defvl}) iteratively to express $u_{L+1}$ 
in terms of the $y_\ell$ we find, for $L=1,2,\ldots,$
\begin{eqnarray}
K_L(\Omega)&=&\ee^{-(L+1)\Omega r}\,\int_0^{W(y_0)}\!\dd y_1 \,\,\ee^{\Omega V(y_1)}
\int_0^{W(y_1)}\!\dd y_2 \,\,\ee^{\Omega V(y_2)}\nonumber\\
&&\phantom{\ee^{-(L+1)\Omega r}\,}\ldots\,  
\int_0^{W(y_{L-1})}\!\dd y_L \,\,\ee^{\Omega V(y_L)}
\label{newintKL}
\end{eqnarray}
where we have abbreviated
\begin{equation}
W(y)=r+y-V(y)
\label{defWy}
\end{equation}
and, by convention, put $y_0=0$. A special case is 
$K_0(\Omega)=\ee^{-\Omega r}$.
We have now transformed the phase space integral for $K_L$
to a problem depending
on the parameter $r$ and the interaction potential $V(y)$.
The original parameters \,$X$,\, $a$,\, and the function $A(\xi)$\, [or,
equivalently,\, $f(t)$\,] no longer appear.

\section{Noninteracting theory: $V=0$}
\label{seclinear}

The noninteracting case is obtained by
setting $V=0$ in the preceding development. 
Strictly mathematically it is not needed to study this case
before passing to the next sections. However, from 
a physical point of view it is highly
desirable to have a good idea of the noninteracting system
before introducing interaction.

For $V=0$ the theory depends on the
single parameter $r$. Correspondingly, all derivatives of $f(t)$ beyond
the first one vanish and $f(t)$ is given by 
\begin{equation}
f(t)=X-f'(0)t
\label{flin}
\end{equation}
We shall denote quantities referring to this linear decay curve by an
index $0$. When combining 
the above expression for $f(t)$ with Eq.\,(\ref{deftau}) 
and the definition $r=\rho\tau$
we find that in this noninteracting case $r$ is given by
\begin{equation}
r_0=\frac{\rho a}{f'(0)}
\end{equation}
which is an instance of Eq.\,(\ref{defaldel}) with $g_0=0.$

{\it Combinatorial problem.}\,\,
Having thus found the parameters of the noninteracting problem,
we have to substitute them in 
the general expression (\ref{newintKL}) for $K_L(\Omega)$.
Imposing as before the initial value $\xi_0=X+a$ we
obtain, after changing to the integration variables $x_\ell=y_\ell/r_0$,
\begin{equation}
K_L(\Omega)=\ee^{-(L+1)\Omega r_0}\,r_0^L\,\int_0^{1+x_0}\dd x_1
\int_0^{1+x_1}\dd x_2 \ldots
\int_0^{1+x_{L-1}}\dd x_L 
\label{intIL}
\end{equation}
where $x_0=0$.
The $L$-fold integral in the above equation, that we shall refer to as
$I_L$, constitutes the heart of the
problem. In terms of the analogy with an $L$ particle system the $x_\ell$
are the particle positions. There is no energy associated with the
allowed configurations $(x_1,\ldots,x_L)$, and $\log I_L$ is the entropy
of the system.

Upon converting to the
integration variables $s_\ell=\ell-x_\ell$, 
where $\ell=1,\ldots,L$, we have
\begin{equation}
I_L=\int_0^1\dd s_1\int_{s_1}^2\dd s_2\ldots\int_{s_{L-1}}^L\dd s_L
\end{equation}
The same integral but with all upper integration limits set equal to
$L+1$ is elementary and equals $(L+1)^{L}/L!$\, It represents the phase
space volume for putting $L$ points on $(0,L+1]$,\,
not counting permutations as distinct. Hence
$I_L=[(L+1)^L/L!\,]\,p_L$, where $p_L$ is the probability that $L$ randomly
chosen points on $(0,L+1]$ are such that, for
$k=1,\ldots,L$, the number $M_k$ of points in the interval
$(L+1-k,L+1]$ is less than $k$.

This may still be rephrased as the following nonelementary 
combinatorial problem. Let $L$
balls be put randomly in $L+1$ numbered vases; then $p_L$ is the
probability that the first $k$ vases contain together
at least $k$ balls, for $k=1,2,\ldots,L$. 

We found no direct way to calculate $p_L$ and
invoke a theorem due to Tak{\'a}cs, of which we adapt the proof to
the present context in Appendix A.
The result is that $I_L=(L+1)^{L-1}/L!$\, 

{\it Persistence exponent.}\,\, Using this in Eq.\,(\ref{intIL}) and
substituting in Eq.\,(\ref{KoKL}) we have
\begin{eqnarray}
K(\Omega)&=&\sum_{L=0}^\infty \ee^{-(L+1)\Omega r_0}\,\,r_0^L\,\frac{(L+1)^{L-1}}{L!}
\nonumber\\
&=&\sum_{L=0}^\infty \ee^{-L(\,\Omega\, r_0-\log r_0 -1)+{\cal O}(\log L)}
\end{eqnarray}
where in the last step we have used Stirling's formula.
It is clear that as $\Omega$ is lowered, $K(\Omega)$ diverges when
$\Omega$ attains a value that we shall call $\Omega_{0}$ and
which is given by
\begin{equation}
\Omega_{0}(r_0)=-\,\frac{1}{r_0}\log\frac{1}{r_0}\,+\,\frac{1}{r_0}
\label{defOmlin}
\end{equation}
Because of Eqs.\,(\ref{defOm}) and (\ref{relthetaom1})
the persistence exponent is
\begin{equation}
\theta_{0}=
\rho\Big(\,\frac{1}{r_0}\log\frac{1}{r_0}\,-\,\frac{1}{r_0}\,+\,1\,\Big)
\label{fintheta}
\end{equation}
Converted to the original variables of the problem this becomes
Eq.\,(\ref{resthetalin}) of the Introduction. 
This exponent
may also be arrived at in ways independent of the recursion
relation formalism of this work ({\it e.g.} with the aid of the method of
Ref.\,\cite{DornicGodreche}, Appendix A), and appears in other contexts
as well ({\it e.g.} the recent work of Bauer {\it et al.} \cite{BGL}).
It will appear again in the next section at the end of a very
different calculation.

\section{Interacting theory: $V>0$}
\label{secnonlinear}

\subsection{Small curvature limit}
\label{seccurvatureappr}

The interacting theory has $V>0$ in Eqs.\,(\ref{newintKL})
and (\ref{defWy}). We will not be able to treat the general case, but
only the one in which the series (\ref{defVy}) for $V(y)$
is dominated by its linear term. 
Curiously enough, although we have to suppose $V$ small
and although our
final results for the exponent $\theta$
will be perturbatively close to the zeroth order
expression (\ref{fintheta}) of the previous section,
the solution {\it method} of the present section is {\it nonperturbative} in
the sense that we 
do not start from the $V=0$ solution, and that in the limit $V\to 0$ 
the method of this section ceases to work.\\

The linear term dominates the series (\ref{defVy}) for $V(y)$
in particular
in the following two limits.

${}$\phantom{i}{\it (i)} $g\to 0$ at fixed $r$, with $\tau g_1=g$ and 
$\tau g_k=o(g)$ for $k=2,3,\ldots ;$

{\it (ii)} $g\to 0$ and $r\to 0$ with a fixed ratio $g/r=c$.\\
\noindent
In both limits the curvature parameter $g$ tends to zero, and we shall refer to
them as instances of a {\it small curvature limit}.
The developments of the next two subsections are common to both limits.
We set $g\equiv \tau g_1$, which, by Eq.\,(\ref{solg2}) and relation (\ref{deff})
between $A$ and $f$, is equivalent to Eq.\,(\ref{defgA}) of the Introduction.
Retaining only the linear term in $V$ we get
\begin{equation}
V(y)=g\,y, \qquad \qquad W(y)=r+(1-g)y
\label{pertexp}
\end{equation}
Hence we have
a theory with two dimensionless parameters, $r$ and $g$;
for $g=0$ the noninteracting theory is recovered.

In the developments that follow the higher order terms, 
sup\-pres\-sed in Eq.\,(\ref{pertexp}),
may be taken into account perturbatively to show that
their effect is negligible to leading order. 
Throughout the present section, the discussion will be only in terms of the
interaction constants $r$ and $g$, that we shall consider as independent
parameters. 
In Section \ref{secapplications} we will return to the original variables of
the problem.

\subsection{Recursion for ${K}(\Omega)$}
\label{secrecursionKom}

If Eq.\,(\ref{pertexp}) is substituted in Eq.\,(\ref{newintKL}), it becomes possible 
to carry the $L$ integrals out recursively for arbitrary $g$, as we
shall show now.
It appears that one needs auxiliary functions
$K_L^{(n)}$ and ${K}^{(n)}$ with $n=0,1,2,\ldots.$ These are defined by
\begin{equation}
{K}^{(n)}(\Omega)=\sum_{L=0}^\infty K_L^{(n)}(\Omega)
\label{defKnom}
\end{equation}
in which for $L=1,2,\ldots$
\begin{eqnarray}
K^{(n)}_L(\Omega)&=&\ee^{-[L+(1-g)^n]\Omega r}\int_0^{r+(1-g)y_0}\!\!\dd
y_1\,\,\ee^{\Omega g y_1}\ldots\nonumber\\
&&\hspace{-15mm}
\int_0^{r+(1-g)y_{L-2}}\!\!\dd y_{L-1}\,\,\ee^{\Omega g y_{L-1}}\,\,
\int_0^{r+(1-g)y_{L-1}}\!\!\dd y_{L}\,\,\ee^{\Omega[1-(1-g)^{n+1}]y_{L}}\,\,
\label{defKnL}
\end{eqnarray}
and where we have the special case
\begin{eqnarray}
K_0^{(n)}(\Omega)&=&\exp[{-\Omega r(1-g{})^n}]\nonumber\\[2mm]
&\equiv&E_n
\label{defEn}
\end{eqnarray}
When Eq.\,(\ref{pertexp}) is substituted in the functions $K$ and $K_L$ of the
preceding section, one sees that $K=K^{(0)}$ and $K_L=K_L^{(0)}$.
Upon carrying out in Eq.\,(\ref{defKnL}) the integral on $y_L$
we find straightforwardly the recursion relation
\begin{equation}
K_L^{(n)}=b_n\,[K_{L-1}^{(n+1)}-K_{L-1}]
\label{recKLn}
\end{equation}
where
\begin{equation}
b_n=\frac{1}{\Omega}\,
\frac{\ee^{-\Omega r(1-g{})^n}}{1-(1-g{})^{n+1}}
\label{defbn}
\end{equation}
Eq.\,(\ref{recKLn}) is
valid for $L=1,2,\ldots$ and $n=0,1,\ldots,$ and must be supplemented 
with the boundary condition (\ref{defEn}) at $L=0$.
Substitution of Eqs.\,(\ref{defEn}) and (\ref{recKLn}) in Eq.\,(\ref{defKnom})
yields for the $K^{(n)}$ the recursion relation
\begin{equation}
{K}^{(n)}=b_n\,[{K}^{(n+1)}-{K}]+E_n
\label{recKom}
\end{equation}
The existence of this recursion relation is the key to the success of
the present method. We remark that for $g=0$ the
coefficients $b_n$ are undefined and the recursion does not exist; hence this
solution method is {\it nonperturbative}.

If we apply (\ref{recKom}) to ${K}={K}^{(0)}$ and
iterate $n$ times, the result is
\begin{eqnarray}
{K}(\Omega)&=&(B_0E_0+B_1E_1+\ldots+B_nE_n)\nonumber\\[2mm]
&&-\,(B_1+B_2+\ldots+B_{n+1}){K}(\Omega)\,+\,B_{n+1}{K}^{(n+1)}(\Omega)
\label{solrecKom}
\end{eqnarray}
where we abbreviated
\begin{equation}
B_0=1, \quad\qquad B_n=\prod_{j=0}^{n-1}b_j \qquad n=1,2,\ldots
\label{defBn}
\end{equation}
We examine now $b_n$, $B_n$, and $E_n$ for $n\to\infty$.
Eqs.\,(\ref{defEn}), (\ref{defbn}), and (\ref{defBn}) show that
in that limit
\begin{equation}
b_\infty=\Omega^{-1}, \qquad  B_n\,\simeq \,\Pi_\infty\,\Omega^{-n}, 
\qquad E_\infty=1
\label{bE}
\end{equation}
where
\begin{equation}
\Pi_\infty\,=\,\prod_{j=0}^\infty\,\frac{\ee^{-\Omega r\,(1-g{})^j}}{1-(1-g{})^{j+1}}
\end{equation}
Hence for $n\to\infty$ we obtain from (\ref{recKom})
an equation for ${K}^{(\infty)}$ with well-defined coefficients. Using
(\ref{bE}) 
one readily finds the solution
\begin{equation}
{K}^{(\infty)}(\Omega)=\frac{E_\infty-b_\infty{K}(\Omega)}{1-b_\infty}
=\frac{\Omega-{K}(\Omega)}{\Omega-1}
\label{solKinfty}
\end{equation}
For $n\to\infty$ we now replace ${K}^{(n+1)}$
in Eq.\,(\ref{solrecKom}) by ${K}^{(\infty)}$ found in
Eq.\,(\ref{solKinfty}). Upon solving for ${K}(\Omega)$ we get
\begin{eqnarray}
{K}(\Omega)&=&\lim_{n\to\infty}\frac
{B_0E_0+B_1E_1+\ldots+B_nE_n + [B_{n+1}/(1-\Omega^{-1})]}
{B_0+B_1+\ldots+B_n+[B_{n+1}/(1-\Omega^{-1})]} \phantom{xxx}
\label{solKomn}\\[2mm]
&=&\,\frac{\sum_{n=0}^\infty B_nE_n}{\sum_{n=0}^\infty B_n},
\qquad |\Omega|>1
\label{solKOm}
\end{eqnarray}
in which $E_n$ is given by Eq.\,(\ref{defEn}) and where from
Eqs.\,(\ref{defBn}) and (\ref{defbn}) we have $B_0=1$ and
\begin{equation}
B_n=\frac{1}{\Omega^n}\,
\exp\Big[-\Omega\,r{}\,\frac{1-(1-g{})^n}{g{}}\Big]\,
\prod_{j=1}^n\frac{1}{1-(1-g{})^j} \qquad n=1,2,\ldots
\label{exprBn}
\end{equation}
Expression (\ref{solKOm})
constitutes the solution of the problem of this work;  the remaining
analysis is needed to extract the persistence exponent $\theta$ from
it. Eq.\,(\ref{solKOm}) holds for the initial condition $\xi_0=X+a$;
without giving the proof we state that for general $\xi_0$ the same
expression (\ref{solKOm}) is obtained except that in the definition (\ref{defEn})
of the $E_n$ one should replace $r$ by $\rho u_1$ and remember that
$f(-u_1)=\xi_0.$ 

By Eq.\,(\ref{relthetaom1}) we have $\theta=-\rho(\Omega_1-1)$, where
$\Omega_1$ is the rightmost nonanalyticity of $K(\Omega)$ in the complex
$\Omega$ plane. We expect the relevant nonanalyticities to be due to
zeros of the denominator of Eq.\,(\ref{solKOm}), for which we shall
introduce the special notation
\begin{equation}
H{}(\Omega;r{},g{})=\sum_{n=0}^\infty B_n
\label{defH}
\end{equation}
In view of the remarks of the preceding paragraph this denominator is
independent of the initial condition $\xi_0$. Obviously its zeros can
occur only for $\Omega<0.$
It is furthermore clear in advance
that for $g{}>0$ the persistence probability must decay
at least as fast as for $g{}=0$, whence
$\theta(r{},g{})\geq\theta(r,0)=
\theta_{0}(r{})$. Consequently we expect that
$\Omega_1\leq\Omega_{0}(r)$, where $\Omega_{0}$ is the
function defined in Eq.\,(\ref{defOmlin}).

\subsection{Analysis of ${K}(\Omega)$}
\label{secanalysisKOm}

We shall evaluate $H(\Omega;r{},g{})$ asymptotically in the two
limits $g{}\to 0$ at fixed $r{}$, and $g{},
r{}\to 0$ at fixed $g{}/r{}$.
In order to prepare for these limits we will transform the sum on $n$
in Eq.\,(\ref{defH}) into a contour integral, to which we shall then apply
the stationary phase method. 

It is first of all necessary to extend the definition of the summand 
$B_n$ to arbitrary complex $n$. To that end we consider the function
\begin{equation}
\Gamma_g{}(z)=\frac{\prod_{j=1}^\infty (1-(1-g{})^j)}
{\prod_{j=1}^\infty (1-(1-g{})^{z-1+j})}
\label{defGammadelta}
\end{equation}
which on the positive integers reduces to 
\begin{equation}
\Gamma_g{}(n)=\prod_{j=1}^{n-1}(1-(1-g{})^j) \qquad n=2,3,\ldots
\end{equation}
and $\Gamma_g(1)=1$.
This function $\Gamma_g(z)$ was introduced 
in 1847 by Heine (see Ref. \cite{GasparRahman}); nowadays 
it is usually
defined \cite{GasparRahman}
with an extra factor $g^{1-z}$ on the RHS of
Eq.\,(\ref{defGammadelta}), and then called the $q$-{\it gamma
function}, where $q=1-g$.
The function $\Gamma_g(z)$ of Eq.\,(\ref{defGammadelta})
has various properties reminiscent 
of the ordinary gamma function. In particular, it has poles for
$z=0,-1,-2,\ldots,$ and the residue $R_m$ in 
$z=-m$ is equal to
\begin{equation}
R_m=(-1)^mg{}^{-1}(1-g{})^{\frac{1}{2}m(m-1)}\frac{1}{\Gamma_g{}(m+1)},
\qquad m=0,1,2,\ldots
\end{equation}
We can now express $H$ of Eq.\,(\ref{defH}) with the $B_n$ of
Eq.\,(\ref{exprBn}) as
\begin{equation}
H{}(\Omega;r{},g{})=\frac{g{}}{2\pi{\text{i}}}\int_C{\text{d}}z\,\,{\text{e}}
^{\,\tilde{h{}}(z,\Omega)} 
\label{intHOm}
\end{equation}
in which 
\begin{eqnarray}
\tilde{h{}}(z,\Omega)&\!=&\!z\log(-\Omega) 
-\Omega\, r{}\,g{}^{-1}[1-(1-g{})^{-z}]\nonumber\\
&&-\frac{1}{2}z(z+1)\log(1-g{}) +\log\Gamma_g{}(z)
\label{defh}
\end{eqnarray}
and where
the integration path encloses the poles of $\Gamma_g(z)$.
Equivalently, we may let this path run 
from $-\infty$ to $0$ below the real
axis, encircle the origin counterclockwise, and run from $0$ back to
$-\infty$ above the real axis.
The poles inside this contour exactly generate the terms of the
series in Eq.\,(\ref{defH}).
A factor $(-1)^n$ coming from $(-\Omega)^n$ cancels against
the $(-1)^n$ from $R_n$.

\subsection{Limit $g{}\to 0$ at fixed $r$}
\label{secdelto0}
If one now scales with $g{}$ according to $\nu=gz$ and writes
\begin{equation}
\tilde{h{}}(z,\Omega)=\frac{1}{g{}}\,h{}(\nu,\Omega;g{})
\end{equation}
then the limiting function $\lim_{g{}\to 0}
h{}(\nu,\Omega;g{})\equiv h{}(\nu,\Omega)$ exists and is equal to
\begin{equation}
h{}(\nu,\Omega)=\nu\log(-\Omega)-\Omega\, r{}\,(1-\ee^\nu)+\frac{1}{2}\nu^2
+\int_0^\nu \!\dd\mu\,\log(1-\ee^{-\mu})
\label{defhbis}
\end{equation}
The poles having become dense, 
this function has a branch cut along the negative real axis in the
complex $\nu$ plane.

{\it Stationary points.\,\,}
In the limit $g\to 0$ we may apply the stationary phase method.
It appears that $h(\nu,\Omega)$ has two stationary points
$\nu_{\pm}(\Omega)$.
There is a critical value $\Omega_c$ such that
for $\Omega>\Omega_c$  the $\nu_{\pm}$ are real and
positive, and for $\Omega<\Omega_c$ they are complex conjugate. At 
$\Omega_c$ we have $\nu_-=\nu_+\equiv\nu_c$. The values
$\Omega_c$ and $\nu_c$ are the solution of
\begin{equation}
h{}_\nu(\nu_c,\Omega_c)=0, \qquad h{}_{\nu\nu}(\nu_c,\Omega_c)=0
\end{equation}
where the indices on $h$ indicate derivatives.
These solutions are easily found and read
\begin{eqnarray}
\Omega_c&=&-\,\frac{1}{r{}}\log\frac{1}{r{}}+\frac{1}{r{}}\nonumber\\[2mm]
\nu_c&=&-\log\,\Big(1+\frac{1}{\log r{}}\Big)
\label{crpoint}
\end{eqnarray}
We see that $\Omega_c(r)=\Omega_{0}(r)$, which establishes the
relation of this nonperturbative calculation with the solution of the
noninteracting theory given in Section \ref{seclinear}.
The analysis can be refined in the vicinity of $(\nu_c,\Omega_c)$.
Upon performing a double Taylor expansion in 
\begin{equation}
\Delta\nu=\nu-\nu_c, \qquad \Delta\Omega=\Omega-\Omega_c
\end{equation}
we find, in obvious notation,
\begin{equation}
h(\nu,\Omega)=h(\nu_c,\Omega_c)+\Delta\Omega \,h_\Omega
+\frac{1}{2}\Delta\Omega^2 \,h_{\Omega\Omega}
+\Delta\nu\Delta\Omega \,h_{\nu\Omega}
+\frac{1}{3!}\Delta\nu^3 \,h_{\nu\nu\nu}+\ldots
\label{expnuOm}
\end{equation}
where all derivatives are evaluated at $(\nu_c,\Omega_c)$,
we have used that $h_\nu=h_{\nu\nu}=0$, and
the dots indicate the remaining third and the higher
order terms. The derivatives that it will be useful to know explicitly are
\begin{equation}
h_{\nu\Omega}=r{}, \quad
h_{\nu\nu\nu}=-\log^2 r{}, \quad
h_{\Omega}=-r{}\,\Big(\log(1+\frac{1}{\log r{}})\Big)
\Big(1+\log r{} \Big)^{-1}
\label{derh}
\end{equation}
The stationary point condition $\partial h/\partial\Delta\nu=0$ applied to
Eq.\,(\ref{expnuOm}) now shows that $\Delta\nu$ has to scale as
$\Delta\Omega^{1/2}$ and we find
\begin{equation}
\nu_{\pm}(\Omega)=\nu_c\mp\frac{1}{\log r{}}(2r{}\,\Delta\Omega)^{1/2}
\label{nupm}
\end{equation}
The stationary point integrations are easily carried out.
For $\Delta\Omega>0$ the relevant stationary point is 
$\nu_-$ and the outcome of the integration is positive.
For $\Delta\Omega<0$ the complex conjugate
points both contribute and the result is
\begin{equation}
H(\Omega;r{},g{})=
2H_0(\Omega;r,g)\cos\Big(\frac{1}{g}
\Big(\frac{8\,h_{\nu\Omega}^3}{9\,h_{\nu\nu\nu}}\Delta\Omega^3 \Big)^{1/2}
+\frac{\pi}{4}\Big)
\label{resH}
\end{equation}
where $H_0(\Omega;r,g)$ is positive 
and where it should be remembered that $h_{\nu\nu\nu}$ and
$\Delta\Omega$ are negative.

{\it Zeros of $H(\Omega;r,g)$.}\,\,
Upon substituting in Eq.\,(\ref{resH}) the explicit expressions 
Eq.\,(\ref{derh}) 
for the derivatives of $h$ we see that 
the function $H$ has zeros for $\Omega=\Omega_j$ with
\begin{equation}
\Omega_j=\Omega_c-\frac{1}{2r{}}\Big(\log\frac{1}{r}\Big)^{2/3}
\Big((4j-1)\frac{3\pi}{4}\,g{}\Big)^{2/3} \qquad j=1,2,\ldots
\label{exprOmj}
\end{equation}
For $j=1$ we obtain the rightmost singularity of ${K}(\Omega)$ in
the complex $\Omega$ plane. Hence by Eq.\,(\ref{relthetaom1}) we obtain
for the persistence exponent $\theta$ the result
\begin{equation}
\theta\,\simeq\,\theta_{0}(r)\,+\,
\rho\frac{1}{2r{}}\Big(\log\frac{1}{r}\Big)^{2/3}
\Big(\frac{9\pi}{4}\,g{}\Big)^{2/3} \qquad r {\mbox{ fixed}},\,g\to 0
\label{thetarg}
\end{equation}
with the function $\theta_{0}$ given by Eq.\,(\ref{fintheta}).
When reconverted to the original variables this gives the result
announced in the Introduction.
The second term on the RHS of Eq.\,(\ref{thetarg}) represents the
leading order
curvature correction to the persistence exponent. It is nonanalytic at
zero curvature.

\subsection{Limit $g{},r{}\to 0$ with fixed ratio $g{}/r{}$}
\label{secaldelto0}

We set $g=cr$. This limit 
requires an independent evaluation starting from 
Eqs.\,(\ref{intHOm}) and (\ref{defh}). The critical point $(\nu_c,\Omega_c)$  
is still given
by Eqs.\,(\ref{crpoint}), but it appears necessary now to scale the deviations
from it as
\begin{equation}
\Delta\bar{\nu}=\log^2\frac{1}{r}\,\Delta\nu, \qquad
\Delta\bar{\Omega}=r\,\log^2\frac{1}{r}\,\Delta\Omega
\label{newscaling}
\end{equation}
Setting for convenience
$\epsilon_r=1/\log(1/r)$ and expanding in $\epsilon_r$ one finds,
instead of Eq.\,(\ref{expnuOm}), the expression 
\begin{equation}
h(\nu,\Omega)=h(\nu_c,\Omega_c)+(\epsilon_r^3+2\epsilon_r^4)\Delta\bar{\Omega}
+\epsilon_r^4\Delta\bar{\Omega}\Delta\bar{\nu}
-\frac{1}{3!}\epsilon_r^4\Delta\bar{\nu}^3
+{\cal O}(\epsilon_r^5)
\label{newh}
\end{equation}
in which $h(\nu_c,\Omega_c)$ is itself of order $\epsilon_r$. The
stationary point condition $\partial h/\partial \Delta\bar{\nu}$ now
leads to $\Delta\bar{\nu}_{\pm}=\mp(2\Delta\bar{\Omega})^{1/2}$, which
when the original scaling is restored is the same as Eq.\,(\ref{nupm}).
The integration through the stationary point involves only the two
${\cal O}(\epsilon_r^4)$ terms in Eq.\,(\ref{newh}).
In view of the proportionality between $g$ and $r$ the
curvature in the stationary point is in this case
of order $(g\log^4g)^{-1}$
instead of $g^{-1}$.
After the calculation is done
the expression for $\theta$ appears to be
exactly what one obtains by naively substituting
$g=cr$ in Eq.\,(\ref{thetarg}), that is,
\begin{equation}
\theta\,\simeq\,\theta_{0}(r)\,+\,
\frac{\rho}{2} \Big(\frac{9\pi}{4}\,c{}\Big)^{2/3}
\frac{1}{r{}^{1/3}}\Big(\log\frac{1}{r}\Big)^{2/3}
\label{thetar}
\end{equation}
Eqs.\,(\ref{thetarg}) and (\ref{thetar}) constitute the main result of this
section. In the following section we shall compare them to direct Monte
Carlo simulations of the process $\xi(t)$. The zeros $\Omega_2,
\Omega_3,\ldots,$ whose explicit expression is furnished by
Eq.\,(\ref{exprOmj}), lead to exponentially small additive corrections
to the leading decay of $Q(T)$.

\section{Examples}
\label{secapplications}

In the following applications we will start from functions $A(\xi)$
defining specific examples of the Langevin-type equation (\ref{defxi}).

{\it First example.}\,\, If in Eq.\,(\ref{defxi}) we take
$A(\xi)=\beta\xi$,
the result is the linear equation 
\begin{equation}
\frac{\dd \xi(t)}{\dd t}=-\beta\xi+a\,\sum_\ell\delta(t-t_\ell)
\label{defxiexp}
\end{equation}
The parameters $r$ and $g$ follow directly from Eqs.\,(\ref{defrA}) and
(\ref{defgA}), respectively, with the result 
\begin{equation}
r=\frac{\rho}{\beta}\,\log(1+\frac{a}{X}), \qquad g=\frac{a}{X+a}
\label{exprrg}
\end{equation}
As $X$ becomes large, $r{}$ and $g{}$ tend to zero
simultaneously with the fixed limiting ratio
\begin{equation}
c=\lim_{X\to\infty}\frac{g{}}{r{}}= \frac{\beta}{\rho}
\label{reldeltaalpha}
\end{equation}
Hence we are in the situation of Subsection \ref{secaldelto0}.
Eq.\,(\ref{deftau}) then leads to
$\tau=\beta^{-1}\log(1+a/X)$,
so that for $X\to\infty$ we have $\tau\to 0$; since $y$ in
Eq.\,(\ref{defVy}) is of order $\tau$, the series for $V(y)$
is one in ascending powers of $\tau$
and we were justified in Subsection \ref{seccurvatureappr}
to neglect the nonlinear terms in $V(y)$. 
If we now substitute expression
(\ref{exprrg}) 
for $r$ and $c$ in Eq.\,(\ref{thetar}) and neglect subleading terms in
the curvature correction, we find the level exponent
$\theta(X)$ given in Eq.\,(\ref{resthetaexp}) of the Introduction.

We have performed a Monte Carlo simulation of the proces $\xi(t)$ of
Eq.\,(\ref{defxiexp}) and determined the persistence exponent
$\theta(X)$ for $X$ ranging from the average $\langle\xi\rangle=\rho
a/\beta$ up to eight times that value. Fig.\,1 shows 
the Monte Carlo data for $\theta(X)$ along with the theoretical result,
Eq.\,(\ref{resthetaexp}),
for asymptotically large $X$. There are no adjustable parameters. The
dashed curve ("free theory") represents only the first term on the RHS
of Eq.\,(\ref{resthetaexp}); the solid curve ("interacting theory", 
full Eq.\,(\ref{resthetaexp})) includes the leading order
curvature correction, which
is the main result of this work. This correction appears to be an important
effect. The excellent agreement between the interacting theory and the
simulation data strongly suggests that higher order corrections to
Eq.\,(\ref{resthetaexp}) vanish for $X\to\infty$. 

\begin{figure}[!h] 
\begin{center}
    \epsfig{width=6.1cm, height=8cm, angle=-90, file=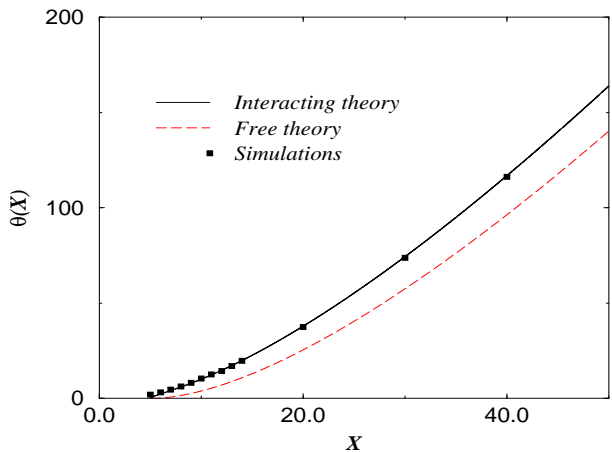}
    \caption{Persistence exponent $\theta$ as a function of the threshold
    $X$ for the process $\xi(t)$ of Eq.\,(\ref{defxiexp}) 
    with $a=1,\, \rho=10,$ and
    $\beta=2$. The average value of this process is
    $\langle\xi\rangle=5.$
    The error bars of the simulation data
    are smaller than the symbols.
    The interacting theory expanded to leading order
    (solid line, Eq.\,(\ref{resthetaexp}) of this
    work) is in
    excellent agreement with the simulations.}
\end{center} 
\end{figure}

\begin{figure}[!h] 
\begin{center}
    \epsfig{width=6.1cm, height=8cm, angle=-90, file=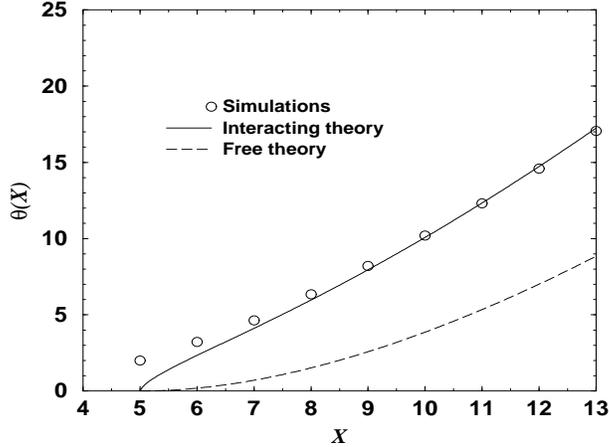}
    \caption{Same as Fig.\,1, zoomed on 
    thresholds $X$ close above the average
    $\langle\xi\rangle=5$.}  
\end{center} 
\end{figure}
Fig.\,2 shows a zoom on values $X\gsim\langle\xi\rangle$;
the leading order behavior of the
interacting theory (solid curve) still represents a considerable
improvement over the free theory, but as $X\to\langle\xi\rangle$,
higher orders in the
expansion become necessary.
For $X<\langle\xi\rangle$
the expansion of this work does not apply.

{\it Second example.}\,\,
Let $A(\xi)$ be such that for some small parameter $\epsilon$ 
\begin{equation}
A(\xi)={\cal A}(\epsilon\xi)
\label{defAeps}
\end{equation}
Eq.\,(\ref{deff}) may then be recast in the form
\begin{equation}
\epsilon t=-\int_{\epsilon X}^{\epsilon f(t)}\frac{\dd x}{{\cal A}(x)}
\label{defepsf}
\end{equation}
whence it follows that $f(t)$ scales as
\begin{equation}
f(t)=\epsilon^{-1}{\cal F}(\epsilon t;\epsilon X)
\label{solepsf}
\end{equation}
We have by construction $f(0)=X$ as before. Furthermore
\begin{equation}
f^{(n)}(0)=\epsilon^{k-1} {\cal F}^{(k)}(0;\epsilon X)
\label{defepsfX}
\end{equation}
where the differentiations of ${\cal F}$ are with respect to
its first argument. If now we agree to choose $X$ of order
$\epsilon^{-1}$, then the $k$th derivative of $f$ is of order
$\epsilon^{k-1}$. This guarantees that $g_k$ is of order $\epsilon^k$,
as is illustrated by Eqs.\,(\ref{solg1}) for $g_1$ and
$g_2$. 
Hence the conditions of limit {\it (i)} are fulfilled and
the calculation of Subsection
\ref{secdelto0} applies. 

In order to test the nonanalytic dependence on the curvature parameter
$g$ in
Eq.\,(\ref{thetarg}) we
have performed a Monte Carlo simulation of $\xi(t)$
for the particular choice
\begin{equation}
f(t)=X-\beta t +\epsilon t^2
\end{equation}
that is,\, ${\cal F}(x,y)=x-\beta y+y^2$. The corresponding $\cal{A}$
follows from Eq.\,(\ref{defepsf}) and by means of Eqs.\,(\ref{defAeps})
and (\ref{defgA}) we find
\begin{equation}
g=1-\Big(1+\frac{4\epsilon a}{\beta^2}\Big)^{-1/2}
\end{equation} 
Monte Carlo simulations were carried out at fixed $a$ and $\beta$ for
various values of $\epsilon.$ In Fig.\,3 we show 
the persistence exponent $\theta$ as a
function of $g$, together with the theoretical $g^{2/3}$ law of
Eq.\,(\ref{thetarg}). The agreement is excellent. 

\begin{figure}[!h] 
\begin{center}
    \epsfig{width=6.1cm, height=8cm, angle=-90, file=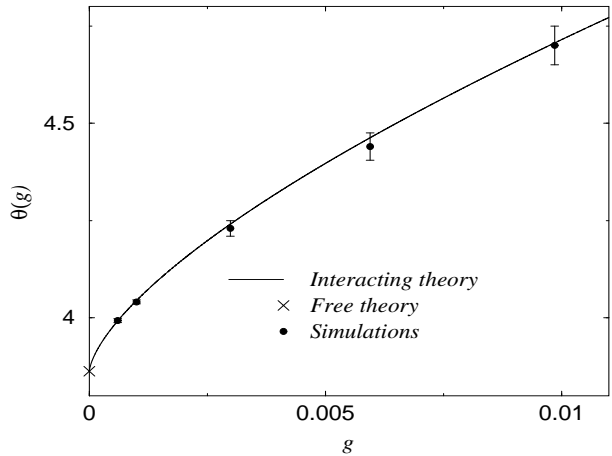}
    \caption{Persistence exponent $\theta$ for fixed threshold $X$
    as a function of the
    curvature parameter $g$ for the second example of Section
    \ref{secapplications} with $a=1,\, \rho=10,$ and $\beta=20.$
    The solid line represents the
    theoretical $g^{2/3}$ law (Eq.\,(\ref{thetarg}) of this work), which appears to
    provide an excellent description of the data.}
\end{center}
\end{figure}

\section{Limit of Gaussian noise}
\label{seccomparison}

The Langevin equation (with white Gaussian noise) and its
extension to
colored Gaussian noise are at the basis of much recent work on
persistence; see {\it e.g.} the recent review by Majumdar
\cite{Majumdar}. 
There is a large body of knowledge today 
about the persistence properties of such Gaussian Markovian
processes, and
a perturbative method 
around the Markovian case has recently
been devised by Majumdar and Sire \cite{MajumdarSire} (see also Oerding
{\it et al.} \cite{OCB} and Majumdar {\it et al.} \cite{SMR}). 
The equation of this work, 
Eq.\,(\ref{defxi}), with jumps of arbitrary finite size $a$,
provides, on the contrary,  an example of strongly non-Gaussian noise.
In this section we show how for $a\to 0$ the Gaussian limit is
approached. This limit, just as the one of zero curvature considered in
Section \ref{seclinear}, 
is a singular point in parameter space.

\subsection{Gaussian persistence exponent $\theta_G$}
\label{secthetaG}

Let $\zeta(t)$ obey the linear Langevin equation
\begin{equation}
\frac{\dd\zeta}{\dd t}=-\beta\zeta + L(t)
\label{defzeta}
\end{equation}
where $L(t)$ is Gaussian white noise of average $\langle L(t)\rangle=0$
and correlation
\begin{equation}
\langle\zeta(t)\zeta(t')\rangle=\Gamma\delta(t-t')
\label{defzz}
\end{equation}
The level exponent $\theta_G(Z)$ for this process, associated with the
probability for $\zeta(t)$ not to have crossed a preestablished threshold
$\zeta=Z$ in a time interval
has not to our knowledge been calculated in the literature. The related
exponent associated  with crossing upward through the threshold has been
considered by Krapivsky and Redner \cite{KrapivskyRedner} (see also
Turban \cite{Turban}). It is easy to
find $\theta_G(Z)$ by a method similar to theirs, as we will show
now. The probability distribution $P(\zeta,t)$ for the process
(\ref{defzeta}) evolves in time \cite{VanKampen} according to the
Fokker--Planck equation
\begin{equation}
\frac{\p P}{\p t}=\beta\frac{\p}{\p \zeta}\zeta P + 
\frac{\Gamma}{2}\frac{\p^2}{\p\zeta^2}P 
\label{FP}
\end{equation}
The persistence exponent $\theta_G$ is the eigenvalue of 
the slowest decaying mode for $Z\!<\!\zeta\!<\!\infty$
satisfing the boundary condition $P(Z,t)=0$. We set $P(\zeta,t) =
P(\zeta)\exp(-\theta_G \,t)$. It is well-known that the equation for
$P(\zeta)$ can be transformed to the eigenvalue equation for the quantum
harmonic oscillator. This fact has been exploited in
previous work \cite{MajumdarSire,SMR,KrapivskyRedner} on persistence
exponents. Here, in view of our interest in the interval $Z<\zeta<\infty$
and the limit of large $Z$, we must transform  
\begin{equation}
\tilde{P}(\mathaccent126{\zeta})=\ee^{\frac{\beta}{2\Gamma} \zeta ^2} P(\zeta) 
\label{defphiy}
\end{equation}
with $\mathaccent126{\zeta}=\lambda ^{1/6}(\sqrt{ 2\beta / \Gamma} \,
\zeta-2\sqrt{\lambda})$ and $\lambda=1/2+\theta_G / \beta$. Then
$\tilde{P}(\mathaccent126{\zeta})$ satisfies the eigenvalue problem 
\begin{equation} 
\left(\frac{\dd^2}{\dd \mathaccent126{\zeta}^2} -
\mathaccent126{\zeta}\left(1+\frac{1}{4}\mathaccent126{\zeta}\lambda^{-2/3}\right)
\right)\tilde{P}(\mathaccent126{\zeta}) = 0, \qquad  \tilde{P}(\mathaccent126{Z}) = 0  
\label{eqnphiy} 
\end{equation}
In the limit of high threshold $Z$ we expect $\theta_G$, and therefore
$\lambda$, to diverge. Hence in this limit
\begin{equation}
\left(\frac{\dd^2}{\dd
\mathaccent126{\zeta}^2}-\mathaccent126{\zeta}\right)\tilde{P}(\mathaccent126{\zeta})
= 0, \qquad  \tilde{P}(\mathaccent126{Z}) = 0 
\label{eqnpsir}
\end{equation}
The solution of Eq.\,(\ref{eqnpsir}) that vanishes for
$\mathaccent126{\zeta} \to \infty$ is the Airy function $\Ai
(\mathaccent126{\zeta})$. The boundary condition $\Ai
(\mathaccent126{Z})=0$ leads to $\mathaccent126{Z}=a_1$, where
$a_1=-2.3381$\ldots\, is the first zero of $\Ai$. This condition fixes
$\theta_G$ in terms of $Z$; upon expanding for large $Z$ one finds 
\begin{equation}
\theta_G (Z)= \frac{\beta^2 Z^2}{2\Gamma} + |a_1| \beta
^{{\frac{2}{3}}}{\left(\frac{\beta^2 Z^2}{2\Gamma}
\right)}^{\frac{1}{3}} + \ldots \qquad (Z \to \infty) 
\label{eqnthetaG} 
\end{equation}
which is the desired result.

\subsection{Gaussian limit}
\label{secglimit}

\subsubsection{Limiting procedure}
\label{seclimproc}

In Eq.\,(\ref{defxi}) we substitute now  $\xi=\zeta + \rho a/\beta$ and
$a\sum_k \delta(t-t_k) - \rho a = L(t)$ and take
the "Gaussian" limit, defined as  
\begin{equation}
\rho \to \infty, \quad a \to 0 \qquad \text{with }
\Gamma=\rho a^2 \text{ fixed}
\label{limgauss}
\end{equation}
The result is that  Eq.\,(\ref{defzeta}) appears. One easily verifies
that $\langle L(t)\rangle=0$ and that the
cumulants of $L$, which  for $n=2,3, \ldots$ are given by 
\begin{equation}
\langle L(t_1)\ldots L(t_n) \rangle _c=\rho a^n \prod_{k=1}^{n-1} \delta(t_k-t_{k+1})
\label{eqncum}
\end{equation}
vanish in the limit of Eq.\,(\ref{limgauss})
when $n\geq3$. Hence $L(t)$
is Gaussian white noise. The above transformation changes the threshold $X$ into $Z = X -
\rho a/\beta$. 
One now expects that the Gaussian persistence exponent $\theta_G(Z)$,
found by direct calculation at the end of the previous subsection,
should also be accessible as a limiting case of our general approach.
Naively, one may attempt to obtain $\theta_G(Z)$ by taking the Gaussian
limit, followed by the limit $Z\to\infty$, in
expression (\ref{thetarg}) for $\theta$.
After a short calculation that procedure leads to 
\begin{equation}
\theta_G(Z) = \frac{\beta^2 Z^2}{2\Gamma} +  { \left
( \frac{9\pi}{8}\right)}^{\frac{2}{3}}\beta
^{{\frac{2}{3}}}{\left(\frac{\beta^2 Z^2}{2\Gamma}
\right)}^{\frac{1}{3}} + \ldots \qquad (Z \to \infty) 
\label{eqnthetalimG}
\end{equation}
This differs from the exact result, Eq.\,(\ref{eqnthetaG}), only by the
numerical value 
of the coefficient of the subleading term; moreover, the difference 
($(9 \pi /8)^{2/3} \simeq 2.3203\ldots$ versus $|a_1|=2.3381\ldots$) is
only about one percent! Nevertheless, Eq.\,(\ref{eqnthetaG}) is right and
(\ref{eqnthetalimG}) is not. The rather obvious reason is that the
Gaussian limit
(which implies $aZ\to 0$), followed by $Z\to\infty$,
does not commute with the limit that was taken to arrive at Eq.\,(\ref{thetarg})
({\it viz.} $r,g \to 0$ at fixed $r/g$, which implies $aZ\to\infty$).
In order to find $\theta_G(Z)$ within the formalism of the preceding
sections it is
necessary to start again from the integral representation of
$H(\Omega;r,g)$ in Eq.\,(\ref{intHOm}).
Below we will see how to do that. 

\subsubsection{Calculation of $\,\theta_G$}
\label{seccalcthetaG}

Let us consider $H(\Omega;r,g)$ of
Eq.\,(\ref{intHOm}). In view of Eqs.\,(\ref{defh})-(\ref{defhbis}) 
it is represented as an integral on $\nu$ of
the function $\exp({g^{-1}h(\nu,\Omega)})$. 
The Gaussian limit is controlled by the parameter $a$, which should
tend to zero.
At fixed $\Gamma=\rho a^2$ and $Z=X-\rho a/\beta$ we find from
Eq.\,(\ref{exprrg}) that in that limit $g=\gamma a^2 +{\cal O} (a^3)$
and $r=1-a\gamma Z +{\cal O}(a^2)$ with
$\gamma=\beta/\Gamma$. 
We recall now Eq.\,(\ref{relthetaom1}), which says that $\theta=\rho(1-\Omega_1)$.
Expecting $\theta$ to approach a finite limit $\theta_G$,
we set $\Omega=1-a^2W$, where $W$ is the appropriately scaled
variable for the relevant region of the complex frequency plane.
Hence, if the rightmost zero of $H(\Omega;r,g)$ in this
plane occurs for $W=W_1$, then
\begin{equation}
\theta_G=\lim_{a\to 0}\, \rho(1-\Omega_1)=\Gamma W_1
\label{relthetaGW1}
\end{equation}

{\it Stationary points.} As a preliminary we consider the stationary
points of $h(\nu,\Omega)$.
Expanding the equation $\p h/\p \nu = 0$ for small $a$ 
while anticipating that 
$\ee^{\nu}$ will be small we find that
these points are solutions of
\begin{equation}
-a^2W + (1-a\gamma Z
)\ee^{\nu} - \ee^{\nu} -
\frac{1}{2}\ee^{2\nu} + \ldots =0 
\label{eqnstatpoints}
\end{equation}
where the dots represent terms of higher order in $a$ and $\ee^{\nu}$.
This shows that there exist solutions with the scaling
$\mbox{Re}\,\nu_{\pm} \sim \log a$ for $a\to 0$. Solving explicitly we obtain
\begin{equation}
\ee^{\nu_{\pm}}=a\,\gamma Z\left(-1 \pm \sqrt{1-\frac{2W}{\gamma^2 Z^2}}\right) 
+ {\cal O}(a^2)
\label{solnupm}
\end{equation}
In the above expression there appears a critical value of $W$
equal to $W_c = \tfrac{1}{2}\gamma^2 Z^2$. 
For $W\!>\!W_c$, which we expect to
be the relevant regime, the stationary points therefore are 
$\nu_{\pm} = -A -\ii \pi \pm\ii \mu^*$ with 
\begin{equation}
-A=\log (a\sqrt{2W})+{\cal O}(a\log a),\qquad \mu^*=\arccos
\frac{\gamma Z}{\sqrt{2W}} + {\cal O}(a)
\label{eqnAmu}
\end{equation}
Instead
of the variable of integration $\nu$ we will henceforth
use $\mu$ defined by
\begin{equation}
\nu = -A -\ii \pi + \ii \mu
\end{equation}
We will not exploit directly, in what follows, our knowledge of $\mu^*$.

{\it Gaussian limit.}
We consider $h(\nu,\Omega)$ of Eq.\,(\ref{defhbis}) 
as a function of $\mu$. After some calculation we find
that for small $a$ 
\begin{equation}
h(\nu,\Omega) = h(-A - \ii \pi,\Omega) + 
a^2\gamma k(\mu,W) + {\cal O}(a^3)
\end{equation}
with
\begin{equation}
\gamma k(\mu,W)=-\ii W\mu + \gamma Z\sqrt{2W}
(\ee^{\ii \mu}-1) -\tfrac{1}{2}W(\ee^{2 \ii \mu}-1) 
\label{exprk}
\end{equation}
In the limit $a\to 0$
the function $H(\Omega;r,g)$ may therefore be rewritten as the integral
\begin{equation}
H(\Omega;r,g)= D \int\dd \mu \,\, \ee^{k(\mu,W)}
\label{intGauss}
\end{equation}
with $k(\mu,W)$ given by Eq.\,(\ref{exprk}) and where $D$ 
diverges when $a$ goes to zero. However, $D$ will divide out 
in Eq.\,(\ref{solKOm}) against the same
factor in the numerator of $K(\Omega)$. 
This completes the Gaussian limit.
There is no small parameter left in the integral in
Eq.\,(\ref{intGauss}).

{\it Limit of large} $Z$.
This integral may be
reduced to a more elementary one in the
limit of large threshold $Z$.
The reason is that then the relevant values of $W$ are close to
$W_c$. We adopt the scaling
\begin{equation}
W\,=\,W_c(1+wZ^{-4/3})\,=\,\tfrac{1}{2}\gamma^2 Z^2(1+wZ^{-4/3})
\label{relWw}
\end{equation}
which will be justified by the results.
We now consider the full Taylor series in $\mu$ of
$k(\mu,W)$. Upon expanding each of its coefficients for
large $Z$ and retaining only the leading term we get
\begin{eqnarray}
k(\mu,W) &=& 
-\tfrac{1}{2}\gamma wZ^{2/3}\ii\mu 
-\tfrac{1}{2}\gamma wZ^{2/3}\dfrac{(\ii\mu)^2}{2!}\nonumber\\ 
&&-\gamma Z^2\,\sum_{n=3}^\infty(2^{n-2}-1)\frac{(\ii\mu)^n}{n!}
\label{eqndevknu} 
\end{eqnarray}
If now the integration variable is scaled
according to
$\mu = \lambda(\gamma Z^2)^{-1/3}$, then in the large $Z$ limit 
all terms in Eq.\,(\ref{eqndevknu}) 
except those with $n=1$ and $n=3$ go to zero.
We are left with
\begin{equation}
H(\Omega;r,g)\,\sim\,\int\dd\lambda\,\,\ee^{-\frac{1}{2}\gamma^{2/3} w\,\ji\lambda
+\frac{1}{6}\ji\lambda^3} 
\qquad(a=0; \,Z\to\infty)
\label{intAi}
\end{equation}
which is the integral representation of the Airy function. The only
dependence left is on the variable $w$.
Let the rightmost zero of $H(\Omega;r,g)$ in the complex frequency plane
occur for $w=w_1$.
We see now that $w_1$ is the solution
of $\Ai(\gamma^{2/3}w_1)=0$, whence
\begin{equation}
w_1 = |a_{1}|\left(2 \Gamma / \beta \right)^{2/3}
\end{equation}
Upon relating $w_1$ to $W_1$ by Eq.\,(\ref{relWw}) and using Eq.\,(\ref{relthetaGW1})
we finally get the expression of Eq.\,(\ref{eqnthetaG}) for $\theta_G$.

{\it Discussion.}\,\,
It is instructive to return to the quantity $\mu^*$ given by
Eq.\,(\ref{eqnAmu}). The two
stationary points are separated by a distance $2\mu^*$, and substituting
the various scaling transformations we see that, as $Z\to\infty$, they
have in terms of
$\lambda$ the finite distance $2\lambda^*=2\gamma^{1/3}w^{1/2}$.
We now observe the mechanism that is at work here.
In Section \ref{secnonlinear}, for $a$ finite, hence
far from the Gaussian limit, 
$H(\Omega;r,g)$ is the sum of contributions from 
two stationary points at infinite separation ($\sim g^{-1}$ with $g\to
0$) in the $\nu$ plane;
as the Gaussian limit is approached, the two stationary points come
within finite distance of one another,
and their contributions cannot be separated any longer.
This "interaction" between the stationary points
leads to the replacement of the cosine in
Eq.\,(\ref{resH}) 
by the Airy function in Eq.\,(\ref{intAi}), and finally affects by about one percent the
coefficient of the subleading term of the persistence exponent.

\section{Conclusion}
\label{secconclusion}
Beside many Gaussian persistence problems, there are also
non-Gaussian ones occurring in statistical physics.
We have pointed out and studied one class of such problems, associated
with the specific non-Gaussian stochastic process that satisfies
Eq.\,(\ref{defxi}). Its relation to several questions in statistical
physics has been indicated in the Introduction.
The sample functions of this
process are deterministic curves
interrupted at
random instants of time by upward jumps. 
Among these, a zeroth order subclass is constituted by
"random sawtooth" functions, characterized by linear decay with
fixed slope.
The persistence exponent $\theta_{0}$
of this subclass is easy to find.
We then perturb around this zeroth order problem by
introducing in the decay 
a small curvature of strength controlled by a parameter $g$. 
As a consequence we have to
deal with what is essentially a one-dimensional
interacting particle system with coupling constant $g$, and
the mathematics becomes considerably more complicated.
The case of greatest importance covered by the present work is
the linear equation, with exponential decay curve,
that prevails for $A(\xi)=\beta\xi$ in
Eq.\,(\ref{defxi}). Our result for this case is an asymptotic
expansion, Eq.\,(\ref{resthetaexp}), of the persistence
exponent $\theta(X)$ in the limit of high threshold $X$.

The same equation for level $X=\langle\xi(t)\rangle$, which is outside
of the domain of the asymptotic expansion of this work, has recently been
considered by Deloubri{\`e}re \cite{Deloubriere}. 
It would be of definite interest to extend
Eq.\,(\ref{defxi}) to {\it random} upward jumps $a_k$ at time $t_k$,
given that specific distributions of jump sizes $a_k$ naturally
occur in several models of statistical physics
\cite{GDH,FreitasLucena}. 

\section*{Acknowledgments}

The authors acknowledge a useful discussion with C.~Sire.
They thank N.M.~Temme for correspondence and for
pointing out the connection of Eq.\,(\ref{defGammadelta}) with
the $q$-gamma function. The relevance of Ref.~\cite{FreitasLucena} as an
example was pointed out to them by Professor L.S.~Lucena. The relevance of
Refs.\,\cite{DornicGodreche,BGL} to the $V=0$ problem was indicated to them
by C.~Godr{\`e}che.

\appendix

\section {A theorem by Tak{\'a}cs}

We consider the problem of determining the probability $p_L$ that occurs
Section \ref{seclinear}. Let the variables $M_k$ be those defined there.
It is natural to set in addition $M_0=0$ and $M_{L+1}=L$, so that 
our problem is to find
\begin{equation}
p_L\,=\,\mbox{Prob}\,\{k-M_k>0 \mbox{ \,for\, } k=1,\ldots,L+1\}
\label{expr2pL}
\end{equation}

Relevant to this problem is {\sc Theorem} 3 by Tak{\'a}cs~\cite{Takacs}, 
which concerns
nondecreasing random functions on line segments. 
The author \cite{Takacs} indicates that 
this theorem has an analog valid for nondecreasing
random sequences. For the present case the full proof runs as follows.

The range of the index $k$ may be extended to arbitrary positive $k$ by
the definition 
\begin{equation}
M_{L+1+k}=L+M_k 
\label{defMk}
\end{equation}
This amounts to repeating the 
set of random point on $0<s\leq L+1$ periodically in the segments 
\,$n(L+1)<s\leq (n+1)(L+1)$,\, where $n=1,2,\ldots.$ The random variable
$M_{k+\ell}-M_k$, where $k=0,1,2,\ldots$ and $\ell=1,2,\ldots,$ 
represents the number of points in the interval $k<s\leq k+\ell$, and
the probability distribution of this variable is obviously independent
of $k$. Let now for $k=0,1,2,\ldots$
\begin{equation}
\delta_k=\left\{
\begin{array}{ll}
1&\mbox{ if\, } M_{k+\ell}-M_k<\ell \mbox{ \,for\, }
\ell=1,2,\ldots\\[2mm]  
0\phantom{xx}&\mbox{ otherwise}
\end{array}
\right.
\label{defdeltak}
\end{equation}
Then the probability distribution of $\delta_k$ does not depend on $k$,
and $\delta_{k+L+1}=\delta_k$. 
It is easy to verify that $M_{k+\ell}-M_k<\ell$
holds for all $\ell$ if it holds for $\ell=1,2,\ldots,L+1$.
Hence Eq.\,(\ref{expr2pL}) shows that
$p_L$ is the pro\-ba\-bi\-lity that $\delta_0$ be equal to 1.
We may write equivalently
$p_L=\langle\delta_0\rangle$, where the average is on all random
sequences $M_1,\ldots,M_L$. But since all $\delta_k$ have the same
distribution, hence the same average, we also have
\begin{equation}
p_L\,=\,\frac{1}{L+1}\sum_{k=1}^{L+1}\langle\delta_k\rangle
\,=\,\frac{1}{L+1}\,\langle\sum_{k=1}^{L+1}\delta_k\rangle
\label{expr3pL}
\end{equation}
We consider now the sum on the $\delta_k$ in the last member of
the above equation. The
condition for $\delta_k$ to equal 1 may be rewritten as
\begin{equation}
j-M_j>k-M_k\quad \mbox{ for all\, } j=k+1,\ldots,k+L+1
\label{conddeltak}
\end{equation}
In the range $k\leq j\leq k+L+1$ the function $j-M_j$ has the initial
value $k-M_k$ and the final value $k+L+1-M_{k+L+1}=k-M_k+1$, where we
used the definition (\ref{defMk}).
If $\delta_k=1$, then $j-M_j\geq k-M_k+1$
for all $j=k+1,\ldots,k+L+1$, and this means that
$\delta_{k+1}=\ldots=\delta_{k+L}=0$. 
Hence $\sum_{j=k}^{k+L}\delta_j$ can be equal only to 0 or to 1.
We now prove that in fact it equals unity.
For it to be zero, all $\delta_j$ in the range of summation would
have to vanish, whence we would have $\delta_j=0$ for all $j\geq k$.
There would then exist an 
increasing sequence $\{j_r\}_{r=0}^\infty$ (where $j_0=k$)
such that the corresponding sequence
$\{j_r-M_{j_r}\}_{r=0}^\infty$ is nonincreasing. 
This however is in contradiction with the fact that 
$j-M_j$
increases by $1$ whenever $j$ is augmented by $L+1$.
It follows that $\sum_{j=k}^{k+L}\delta_k=1$,\, 
whence by Eq.\,(\ref{expr3pL})
we obtain $p_L=1/(L+1)$ and $I_L=(L+1)^{L-1}/L!$


\end{document}